\pdfoutput=1
\documentclass{article}
\usepackage[utf8]{inputenc}
\usepackage{amsmath,amssymb,amsthm,amsfonts,bm,caption,centernot,color,colortbl,csquotes,graphicx,mathrsfs,mathtools,physics,setspace,subcaption}
\usepackage[sorting=none]{biblatex}
\usepackage{makecell}
\usepackage{smartdiagram}
\usepackage{listings}
\usepackage{geometry}
\usepackage{tikz}
\usetikzlibrary{shapes,arrows}
\theoremstyle{definition}

\newtheorem{example}{Example}[section]
\theoremstyle{plain}

\definecolor{Gray}{gray}{0.9}
\usepackage{authblk}
\begin{document}

\title{A New String Edit Distance and Applications}


\author[$\dagger,\mathsection$]{Taylor Petty}
\author[$\dagger, \ddag$]{Jan Hannig}
\author[$\star$]{Tunde I Huszar}
\author[$\ddag$]{Hari Iyer}

\affil[$\dagger$]{Department of Statistics and Operations Research, The University of North Carolina at Chapel Hill, Chapel Hill, NC, USA 27599}
\affil[$\ddag$]{Statistical Engineering Division, Information Technology Laboratory, National Institute of Standards and Technology, Gaithersburg, MD, USA 20899}
\affil[$\star$]{National Institute of Standards and Technology, Gaithersburg, MD, USA 20899}
\affil[$\mathsection$]{Corresponding author. taylor.petty@unc.edu.}
\date{}

\maketitle

\begin{abstract} 
String edit distances have been used for decades in applications ranging from spelling correction and web search suggestions to DNA analysis. Most string edit distances are variations of the Levenshtein distance and consider only single-character edits. In forensic applications polymorphic genetic markers such as short tandem repeats (STRs) are used. At these repetitive motifs the DNA copying errors consist of more than just single base differences. More often the phenomenon of ``stutter'' is observed, where the number of repeated units differs (by whole units) from the template. To adapt the Levenshtein distance to be suitable for forensic applications where DNA sequence similarity is of interest, a generalized string edit distance is defined that accommodates the addition or deletion of whole motifs in addition to single-nucleotide edits. A dynamic programming implementation is developed for computing this distance between sequences. The novelty of this algorithm is in handling the complex interactions that arise between multiple- and single-character edits. Forensic examples illustrate the purpose and use of the Restricted Forensic Levenshtein (RFL) distance measure, but applications extend to sequence alignment and string similarity in other biological areas, as well as dynamic programming algorithms more broadly. 


\end{abstract}

\section{Introduction}

String similarity is an important tool for many applications, including web search completion suggestions \cite{dynamicquerysuggestions}, spelling correction \cite{compspelling1967}, and ontology alignment \cite{ontologyalignment}. Some of the tools developed in the field of string similarity carry over well to DNA applications, such as DNA-protein matching, sequence assembly, and local similarity searches \cite{chang_lawler_1994}, as well as Longest Common Substring and Longest Common Subsequence algorithms \cite{stringmatchmethods}. Determining similarity of DNA strands is important in phylogenetics, and the volume of available data continues to increase over time. Alignment-free methods have been developed, in part to account for issues that arise when traditional sequence alignment is pushed to its limit. These include the use of geometric representations in various dimensions, as well as graph theory \cite{graphtheorydnasimilarity}.

Most string edit distances consider only single character edits and are variations of Levenshtein distance. These standard string edit distances are inadequate for analyzing similarity between DNA sequences in forensic applications where the goal is to identify the individual who is the source of the DNA in a crime sample. To fill this gap, a new string edit distance is defined, referred to as Restricted Forensic Levenshtein (RFL) distance, that accommodates the addition or deletion of one or more motifs (consecutively repeating sequences of nucleotides 2 to 6 base pairs long) as edit types separate from single-nucleotide insertion, deletion, and substitution. This makes the problem of computing the distance much more difficult. To address this, a dynamic programming algorithm for computing the RFL distance between sequences has been developed. Forensic examples are used to illustrate the purpose and use of RFL distance but the applications extend to sequence alignment and string similarity in other biological applications. The algorithm's novel contribution is in efficiently handling the complex interactions that arise between multiple- and single-character edits.

\subsection{Background}

DNA-based human identification is an extremely powerful tool with many important applications such as identification of perpetrators of crimes, disaster victim identification, DNA typing of skeletal remains, and relationship testing \cite{butler_2015, clayton_whitaker_maguire_1995, skeletal_remains_2001}. The same methods are also applicable in food authenticity, poaching, and counter-bioterrorism \cite{reachwithmps}. DNA measurement technologies have made great strides recently, and now even minute quantities of DNA can be reliably measured \cite{reachwithmps}.

In forensic DNA testing, genetic material present at a crime scene is extracted and amplified through the polymerase chain reaction (PCR). In principle, this technology replicates specific regions of the human genome present in the sample by a repeated process of doubling. Errors may get introduced to the products during the PCR process, and these error sequences are called \textit{artifacts}. Traditionally, analysis of the amplified DNA products is carried out via a length-based measurement using capillary electrophoresis (CE); however, with the introduction of sequencing to the workflow, the amplified products now enter the process of massively parallel sequencing (MPS) providing a sequencing-based measurement for an allele. After the sequencing is complete, the generated data can be used as an input for bioinformatics pipelines. The data analysis provides details about the measured quantity of the expected PCR product sequences (here referred to as \textit{parent} alleles) and numerous artifact sequences related to the range of genetic markers examined in these assays. In typing a standard single-source sample for a set of autosomal markers, the sequences observed should include one to two different parent alleles (depending on whether the sample was homozygous or heterozygous at the autosomal locus) and numerous artifact sequences that are present because of copying errors in the amplification and sequencing process. Each detected sequence is quantified by the number of sequences observed and is commonly referred to as the \textit{depth of coverage} (DoC). Along with the parent alleles there are artifacts that are also detected and are often present at a low DoC, and therefore easily distinguishable from the parent alleles. Each of the artifacts is an error sequence of a true parent, and one can generally infer which parent it came from, but occasionally the parent sequences are similar enough that it is difficult to tell which is the actual parent sequence. For single-contributor DNA samples, this is typically not an issue, but crime scene samples motivating this work are rarely from a single individual. Crime scene samples often consist of multiple individuals, contributing different amounts of genetic information. For example, in a complex mixture (e.\,g. three contributors at 70\%-25\%-5\%) it can be challenging to determine the respective parent allele of each contributor and their corresponding artifact(s).

In this light, quantifying what it means for nucleotide sequences to be {\em similar} becomes a crucial task, which has led to the notion of a generalized Levenshtein distance that allows gains and losses of multiple letters at once (here the change of multiple bases at once in the sequence string), and the development of an algorithm to calculate it. A major contribution of the proposed algorithm is in handling the exponential explosion of complexity when multiple-character edits and single-character edits interact, which requires careful techniques to resolve.

\subsection{Short Tandem Repeats}

The DNA identification community commonly uses short tandem repeats (STRs) as their genetic marker of choice \cite{variationinstr_1994, rapid_efficient_1994}. STR typing methods are accurate and sensitive \cite{dnatyping_str_1993}. The work in this paper is designed for the analysis of DNA samples in forensic applications, but the methods presented here can be applied in other areas that use STR analysis. An STR is a region of DNA that contains consecutive repeats of a subsequence of nucleotides 2-6 base pairs long, called a \textit{motif}. To the left and right of this repeating region is a non-repeating section of DNA called a \textit{flanking region.} For example:

\begin{center}
    ACTCC ATG ATG ATG ATG GGTTCTGA
\end{center}

\noindent Here the motif ATG is repeated four times, which is represented in a short format as [ATG]4.

As sequencing technology has developed, individual nucleotides within STR regions in the human genome can be measured increasingly economically and accurately, whereas in the past people were limited to measuring the total length of the targeted sequence. The standard length-based CE methods do not allow for the detection of intra-motif variations. For a real-data example, consider a sample with a two-allele locus with repeat regions $$\text{A: [TCTA]8 [TCTG]1 [TCTA]1}$$ and $$\text{B: [TCTA]10.}$$ Note A and B are identical by length. When both alleles at a locus share the same length but differ by sequence, they are known as \textit{isometric heterozygotes}, and CE methods cannot discriminate between them. CE types both of these alleles as 10, an allele with a length of ten repeats, and classifies this sample as homozygous at this locus. By recognizing variation at the sequence level, the power to discriminate between individuals increases \cite{nist2016paper}. If an artifact sequence C = [TCTA]7 [TCTG]1 [TCTA]1 was detected in the sequencing output, it would be reasonable to infer that there is a higher likelihood of it being an error derived from A rather than B, since fewer changes are required to edit A into C than B into C. Consequently, it would make sense to infer that the artifact is related to A. With CE, A and B are viewed as identical, and no such conclusions about the artifact can be made. As crime scene samples are most often mixtures, the additional resolution of the alleles and artifacts provided by the use of sequencing can be advantageous compared to solely length-based analysis.

In STR regions, motifs often expand and contract as a unit. This phenomenon of slippage of the polymerase on the template is referred to as \textit{stutter}. The artifacts of \textit{in vitro} stutter products are consistently present in measurable quantity and therefore are accounted for in routine forensic DNA analysis \cite{brookes_bright_harbison_buckleton_2012}. Furthermore, stutter occurs \textit{in vivo} in STR regions even at conservative estimates at least 3-4 orders of magnitude more often than other random point mutations \cite{invitrostutter}. To reflect this, a generalization of string edit distance was used to capture similarity between sequences in MPS output. The goal was to include the addition or deletion of a motif as an edit type with a cost that can vary independently from single-nucleotide edits, to reflect the fact that these motif losses and expansions (called \textit{backward} and \textit{forward stutter}, respectively) occur consistently across various \textit{in vitro} applications \cite{stutterformation}. The proposed algorithm is also applicable to \textit{in vivo} methods such as rare disease diagnosis and cancer progression tracking that use STR markers including their stutter \cite{stuttermodel}.

It is desirable to develop a distance that gives high-frequency artifacts a lower distance to the true alleles than low-frequency artifacts, so it is important to capture both reverse and forward stutter as low-cost events. Standard string edit distances such as the Levenshtein distance modify one character at a time, so the cost of dropping or adding a motif of multiple characters is necessarily equal to the sum of the costs of dropping or adding the individual characters. Furthermore, multiple motifs can stutter within a single STR region, with each motif having a potentially unique cost. Beyond the issue of edit cost, stutters and single-character edits can interact in complicated ways, and algorithms like the Levenshtein distance and its relatives do not begin to address the issues that arise in that interaction.

String edit distances have been in use for decades, with many papers published on different applications. Review of the literature revealed diverse uses, such as a spelling correction algorithm using a generalized metric that allows for generic word-to-word edits \cite{brill_moore_2000}. The aim and methods described in the study are different than those presented here; however, if the code and details for implementation were included in \cite{brill_moore_2000}, direct comparison with the independent method presented here would be of interest.

The ideas behind the RFL distance are applicable to any situation requiring a minimal edit distance solved via dynamic programming, notable among which is graph edit distance \cite{grapheditdistsurvey}. By extending the idea of precomputed motifs to other dynamic programming settings, one could consider a minimal graph distance that allows the addition or loss of an entire subgraph as its own edit to have an individual cost. Therefore, this idea could have implications wherever graph edit distances are used, such as handwriting recognition \cite{handwritingrec}, fingerprint recognition \cite{fingerprint}, and cheminformatics \cite{cheminformatics}.

The proposed algorithm to compute RFL distance has fully customizable costs, and is therefore capable of reflecting the idea that lower costs correspond to higher probability edits. This connects to the notion that more frequent events would be favored in a maximum likelihood approach over lower probability events. 
The Python code for the RFL algorithm is available on GitHub \cite{mygithub}.

\section{Restricted Forensic Levenshtein Distance}


The proposed algorithm handles both single- and multiple-character edits, as well as their interactions. The single-character part of the algorithm is built on the original Levenshtein distance, which will be reviewed below. An introduction of the RFL distance follows.

\subsection{Levenshtein Distance Overview}

String edit distances are based on the minimal number of operations required to transform one string into another. This gives a way to measure dissimilarity of objects in the non-Euclidean space of strings, using the notion that two strings that are similar should differ in only a few characters, while two strings that are dissimilar would require many changes to turn one into the other. The edit distance proposed here could be viewed as a generalization of the Levenshtein distance \cite{Lev65}, which counts the minimal number of single-character deletions, insertions, and substitutions required to transform one string into another. For example, the distance from CAT to CGT is one, from CAT to CA is one, and from CAT to CATTG is two. Other edit distances exist, with different edits allowed. Damerau-Levenshtein considers indels, substitutions, and transpositions \cite{damlev}, while the longest common subsequence metric only looks at indels \cite{longcomsub}. Rather than having a common cost of $1$,  substitution, insertion, deletions, etc. can each have their own unique cost \cite{wagnerlowrance1975}.

The calculation of minimal edit distance is typically done via dynamic programming \cite{LevDynamicProg}. The standard dynamic programming solution to Levenshtein will be briefly discussed below because it is crucial to understanding the novel contributions of the proposed algorithm. Examples will be presented to maximize clarity.

\subsection{Forensic Distance Overview}

The STR regions of DNA that are currently used for forensic identification were historically selected based on the polymorphic length variations of these markers observed in the population. Most of these are tetranucleotide repeats, but di-, tri-, penta- or hexanucleotide repeats also exist in the tandem orientated arrays. At sequence level a trinucleotide marker can be described as a string such as ACTACTACTACT, or abbreviated, as in [ACT]4. Backward stutter refers to losing a motif, or [ACT]3 in this example. Less-common forward stutter refers to gaining a motif, or [ACT]5 here. There are other possibilities, such as substitution, e.g., [ACT]4 to changing into an [ACT]2 AGT [ACT]1 via a C $\to$ G substitution, but these are less common, and already taken into account under the classical Levenshtein distance.

Stutter is more common than insertion or deletion, so it is modeled as a separate edit type. This enables the cost of stutter in the edit distance to be lowered to reflect its higher probability, instead of modeling it as the sum of individual nucleotide changes. In order to accomplish this, the RFL algorithm includes a step of looking at a set of substrings of various lengths with their associated costs from a precomputed dictionary. Whereas the standard dynamic programming solution looks back a single letter, our proposed approach looks back several letters. The particular dictionary of substrings is what gives the RFL algorithm the title ``restricted,'' since the size of the dictionary limits how far back the algorithm looks.

The complexity comes from the interaction of multi- and single-character edits. For example, the sequence [ACT]4 could change to [ACT]5 via forward stutter, then a C could be deleted, for a final result of [ACT]4 AT. This final string could have been attained by inserting the letters A and T from the original [ACT]4, but another option was the forward stutter and single nucleotide deletion. Different edits contribute different (customizable) costs, so one route may be more optimal than the other depending on the parameters. Extending these interactions across multiple edits becomes exponentially complex.

A data-driven approach requires knowing which motifs at each locus stutter at significant rates. For the design of the RFL algorithm, sequence analysis data from 22 autosomal STRs was used from a set of 661 samples from four U.S. populations. The sequence data was previously generated using the PowerSeq Auto/Y System kit (Promega Corp., Madison, WI) and was analyzed using STRait Razor v3.0 \cite{previousSTRanalysis}. The output files were used to identify motifs of length 2-6 that were the basis of observed stutter artifacts at each autosomal locus. This analysis is described in Section \ref{subsec:motifselec}.

\subsection{Details of the RFL algorithm}

The foundation of the RFL algorithm is built on the Levenshtein distance. In this section a perspective on the standard dynamic programming algorithm for classical Levenshtein is discussed, then Weighted Levenshtein is discussed, and lastly the novel modifications are presented. Zero-based indexing for strings will be used, following the conventions of Python, with $s[i:j]$ to represent the substring of $s$ from the $i$-th to the $(j-1)$th character, inclusive. Similarly, $s[i:]$ will represent the substring composed of the $i$th character up to and including the last character, and $s[:j]$ will represent the first $j$ characters, up to index $j-1$, inclusive.

\subsubsection{A Perspective on Dynamic Programming}

From this point on, a working knowledge of the Levenshtein distance is assumed, with the following perspective. To compute the distance between strings $a$ and $b$ with unit edit costs, the base case is that one of the strings is empty. If $a$ is the empty string then insert each character of $b,$ and the number of those insertions is the cost. If $b$ is the empty string, delete each character of $a,$ and the number of those deletions is the cost. If the first characters of $a$ and $b$ are the same, then delete $a[0]$ and $b[0]$ for no cost and the overall distance will be the same as the Levenshtein distance between the remainder of the strings $a[1:]$ and $b[1:]$ (denoted as tail($a$) and tail($b$), respectively).

In case the first characters of $a$ and $b$ don't match, the standard dynamic programming algorithm for Levenshtein distance effectively aligns the leading character of $a$ with the leading character of $b$, then modifies the leading character of $a$ to match the leading character of $b$. Once the leading characters of the strings match, the matching leading characters can be deleted \textit{for free}.

For example, in order to compute the Levenshtein distance from AGTCT to GACT, the A must be deleted and the first T switched to an A, for a cost of two. This eyeball calculation only works for simple examples, so Figure \ref{howstackingworks} illustrates this perspective on the classical Levenshtein algorithm.

\tikzstyle{algo} = [rectangle, draw, fill=blue!20, 
    text width=5em, text centered, rounded corners, minimum height=4em]
\tikzstyle{line} = [draw, -latex']
\tikzstyle{seq} = [draw, ellipse, fill=red!20, node distance=3cm,
    minimum height=2em,align=left]

\begin{figure}
\begin{center}
\begin{tikzpicture}[node distance = 2cm, auto]
    \node [seq] (seq1) {AGTCT\\ GACT};
    \node [algo, below of=seq1] (algo1) {Delete A, top row \\ (cost 1)};
    \node [seq, right of=seq1]  (seq2) {GTCT\\ GACT};
    \node [algo, below of=seq2] (algo2) {Delete both Gs\\ (cost 0)};
    \node [seq, right of=seq2]  (seq3) {TCT\\ ACT};
    \node [algo, below of=seq3] (algo3) {T $\to$ A, top row\\ (cost 1)};
    \node [seq, right of=seq3]  (seq4) {ACT\\ ACT};
    \node [algo, below of=seq4] (algo4) {Delete all\\ (cost 0)};
    \path [line] (seq1) -- (algo1);
    \path [line] (algo1) -- (seq2);
    \path [line] (seq2) -- (algo2);
    \path [line] (algo2) -- (seq3);
    \path [line] (seq3) -- (algo3);
    \path [line] (algo3) -- (seq4);
    \path [line] (seq4) -- (algo4);
\end{tikzpicture}
\caption{The steps to compute the Levenshtein distance from AGTCT to GACT.}
\label{howstackingworks}
\end{center}
\end{figure}
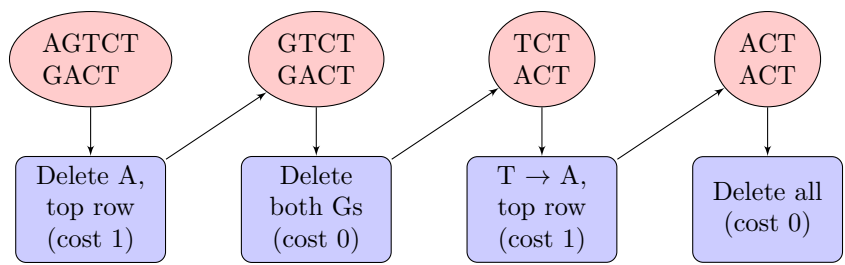

Note that in Figure \ref{howstackingworks} there could have been an additional T concatenated to the beginning of AGTCT and then immediately deleted. This would have been a valid edit, but not a minimal one. At each step the minimal cost must be taken, and this will be accomplished via dynamic programming. For a $n_a$-character string $a$ and an $n_b$-character string $b$, the dynamic programming algorithm constructs an $(n_a+1)\times (n_b+1)$ matrix, and overall it is $O(n_a n_b)$ in time complexity.

\subsubsection{A Note on Weighted Edit Distance}

In a setting where different edits happen with different frequencies, one might want to weight edits differently from each other. This could be because DNA exhibits higher rates of changing a particular character to another, or in spelling correction because certain letters are closer to others on the keyboard. For example, inserting an A could be penalized more than any other letter, and substituting C for the letter G could be incentivized. This could be done by raising the insertion cost of an A to 2 and lowering the substitution cost of G $\to$ C to 0.5. The edit distance algorithm remains the same, but whenever those edits are performed, the cost is different. This might change the optimal edit path from one string to another. Normally, converting CAT into TGT would be two substitutions: C $\to$ T and A $\to$ G. In the case where all substitution costs are equal to 4 and insertion and deletion costs are still equal to 1, the previous edit path involving two substitutions would cost a total of 8, and a new minimal-cost path would be CAT $\to$ AT $\to$ T $\to$ GT $\to$ TGT, which would cost 4. This could be made much more complicated by changing every possible edit cost to be different than every other. The dynamic programming algorithm solution is able to take any possible assortment of costs and compute the minimal cost given those parameters. 

This can easily make the distance non-commutative, i.e., it is a {\em directed} distance. In a case where the substitution cost of C $\to$ G is 3, but all other costs remain equal to 1, then the distance from CAT to GAT is 2 (delete C, insert G; this is still less costly than changing C to G directly), but the distance from GAT to CAT is 1 (changing G to C still costs 1). Therefore the distance hereafter is referred to as being \textit{from} a sequence $s$ \textit{to} another sequence $t$, which represents that the edits begin at $s$ and change into $t.$ A library for the weighted Levenshtein, optimized in Cython, is available \cite{wlev_pypi}.

\subsubsection{Restricted Forensic Levenshtein Distance}

Here the full steps of the RFL algorithm are presented. To compute the distance between strings $a$ and $b,$ if $a$ is the empty string, simply insert each character of $b,$ and the number of those insertions is the cost. If $b$ is the empty string, delete each character of $a,$ and the number of those deletions is the cost. If the first characters of $a$ and $b$ are the same, then delete $a[0]$ and $b[0]$ for no cost and the overall distance will be the same as the Levenshtein distance between the remainder of the strings $a[1:]$ and $b[1:]$ (denoted as tail($a$) and tail($b$), respectively).

In case the first characters of $a$ and $b$ don't match, line the two strings on top of each other with $a[0]$ aligned to $b[0],$ with all edits being made to the parent string $a$ (this becomes important later, when the distance becomes asymmetric). The goal of the RFL algorithm is to make the leading characters of the strings match, in which case the matching leading characters can be deleted \textit{for free.} For instance, the distance from CAT to CGT is the same as from AT to GT, since the leading C does not require editing. Returning to the aligned strings $a$ and $b,$ there are three options. The first is that the leading character of $a$ can be changed into the leading character of $b$, which corresponds to substitution errors in PCR. A new leading character of $a$ can be inserted to match the leading character of $b$, which corresponds to insertion. Lastly, the leading character of $a$ can be deleted so that the new leading character of $a$ matches the leading character of $b$, which corresponds to deletion.

The goal is to be able to include the deletion or insertion of multiple consecutive letters as a separate edit type at a reduced cost. The classical Levenshtein solution is capable of gaining or losing multiple letters, but the cost will always be the sum total of the costs of the necessary single-character interim operations.

The classical Levenshtein algorithm looks at a character and determines what operation at that position gives the lowest cost to change a single letter at that spot when changing one string into another. It does this iteratively, breaking the original problem into sub-problems by addressing a single letter at a time. The RFL modification further looks at multiple characters at once. This will correspond to looking back several positions within the dynamic programming matrix. Only particular substrings will be allowed as their own edit types, as opposed to any general substring. The RFL solution is to create a dictionary of multi-character substrings and their associated insertion and deletion costs. When a substring -- for example CTG -- needs to be inserted or deleted, the relevant cost associated with that substring in the dictionary is retrieved, and that option is included for the algorithm to consider when taking a minimum cost step at that index. This substring CTG corresponds to looking back three steps in the dynamic program.

For the sake of illustration, assume unit costs for all edits, and consider AAAGA, the motif at the Penta D locus. If AAAGA is permitted to be added or deleted all at once, then inserting AAAA will be cheaper if AAAGA is first built and then the G deleted. Inserting AAA would cost the same either way: inserting 3 As, or inserting AAAGA (+1) and then deleting an A (+1) and deleting a G (+1). Inserting AAAGTA would cost 2 (+1 to stutter and +1 to insert), as opposed to costing 6 by inserting one at a time. The RFL algorithm needs to account for this, or else it would give the string AAAGTA a cost of 6 because the motif AAAGA is ``broken'' by the T.

The question naturally arises: if costs are being precomputed for different substrings, which substrings should be included? Suppose that for a specific motif (e.\,g., AAAGA), all possible substrings and their associated costs were included. Note that \textit{any} string with the letters AAAGA (in that order) would be cheaper to insert by first stuttering forward and then adding letters in between, instead of adding them one by one. This is shown in Example \ref{alwayscheaper}.

\begin{example}\label{alwayscheaper}
To build the string
\begin{center}
A(TTTT)A(TTTT)A(TTTT)G(TTTT)A    
\end{center} using edit operations starting with an empty string, 21 letters would need to be inserted, for a cost of 21. This would cost 4 less overall by inserting the motif AAAGA first via stutter (cost of 1) and then inserting all the remaining nucleotides in between (cost of 16). This is true for an unboundedly long string, with any number of characters in between the motif characters as long as the letters AAAGA are present in that order.
\end{example}

Even if a string had the letters A...A...A...G...A with 2,000 nucleotides in between each letter, by including it in the dictionary, the algorithm would qualify it as a modified stutter motif. Thus, not only is an infinite (or virtually infinite) dictionary impossible, it is undesirable and potentially absurd. In addition, there is no evidence to suggest that PCR results in such drastic numbers of errors. The faithfulness of the polymerase used in PCRs for STR typing has improved enough that these only err sporadically, which is supported by the fact that only small deviations were observed in artifacts in this study. Such assignment of stutter cost over excessive length of the string is undesirable, therefore a limit to the extent of the modification was incorporated.

The supplementary materials contain more detailed discussion regarding the complex cases that arise when considering this additional edit type, but a few examples are included here. As mentioned previously, it is possible to have intra-motif substitutions and indels. The STR [ACT]4 could give rise to [ACT]2 AT [ACT]1. The difference in the number of ATG motifs here is $4 - 3 = 1,$ but there was no stutter -- just deletion of the letter C. Including the interaction of stutter with other edit types quickly explodes complexity in the minimal cost computation.
Thus, for every possible nucleotide sequence of length $\leq 2k-1$, where $k$ is the length of the motif, the classical, weighted Levenshtein distances of both inserting these strings from scratch or deleting them are computed, with the requirement that the edit path contains exactly one stutter. Specifically, the precomputed insertions contain short sequences of letters with an associated cost, assuming the edit path \textit{begins with a motif insertion.} The precomputed deletions contain the same strings as the insertion list, but with the cost of deleting, assuming the edit path \textit{ends with a motif deletion.} This data is stored in a pair of dictionaries. In the GitHub resource, this function is called \texttt{lsdp}, an acronym standing for Levenshtein Stutter Dictionary Pair. Specifics are illustrated in Example \ref{example:lsdp}.

\begin{example}\label{example:lsdp}
For the motif TCTA, the stutter dictionary $d$ has two keys: `insert cost' and `delete cost'. Each of these keys gives a dictionary $d[key]$ of substrings and their associated costs. A selection is given as follows, where all edit costs equal to one -- note that the insert and delete dictionaries will always contain the same entries, but different entries are shown here to demonstrate different costs:
\begin{align*}
    d = \{&\text{insert cost: }\\
    &\qquad \{\text{TCTA: } 1,\\
    &\qquad \text{TCTAT: } 2,\\
    &\qquad\text{A: } 4,\\
    &\qquad\text{GGGGGGG: } 8,\\
    &\qquad...\},\\
    &\text{delete cost: }\\
    &\qquad \{\text{TCTATGG: } 4,\\
    &\qquad \text{TCT: } 2,\\
    &\qquad \text{TCTA: } 1,\\
    &\qquad...\}\\
    &\}\\
\end{align*}

Note that strings like `GGGGGGG' and `A' cost more to insert in this dictionary than they do in classical Levenshtein, since the RFL algorithm first inserts a motif for a cost of 1 before modifying it into the desired string. A symmetric fact holds for deletion entries. However, since RFL includes classical Levenshtein operations and takes the lowest-cost path, inserting the string `A' will be considered both as a cost 1 edit (from the classical Levenshtein solution) and a cost 4 edit (from the dictionary), with the lower cost option being taken. 
\end{example}

In the RFL algorithm, all edits are fully customizable. Every edit from one nucleotide to another can be set with a unique cost, and every relevant motif can stutter forward and backward with unique costs.

The classical Levenshtein distance separately considers the cost of an insertion, deletion, or substitution, and takes the minimum of the three at each step to expand the matrix via dynamic programming. To account for all the substrings accessible via stutter, the RFL algorithm considers the same possibilities as classical Levenshtein, and additionally considers at each step the options of inserting or deleting any $l$-character portion, where $l$ ranges from $1$ to $2k-1,$ and where the respective cost is defined by the pre-built dictionary.

\subsubsection{How Restricted To Be?}

Stopping at string length $2k-1$ is a parameter easily changed. Previously discussed was the motivation to consider only motifs that are slightly modified. Insertions were so rare in the data analysed in this study that by the time an additional $k$ individual nucleotides have been inserted into a motif of length $k$, it is desirable for the proposed algorithm to be looking at a potential pair of motifs instead of one highly mutated motif. If a string is gaining or losing $2k$ characters at once, it is more likely due to two separate stutters than to a single stutter that was then affected by an additional $k$ inserts or deletes. Furthermore, in this data set, stutter and substitutions are much more common than deletions, and all are more common than insertions, giving more reason to cap the string length.

These lists of strings and their associated costs are computationally intensive to create (although still generally under a minute), but once one is made (stored as a dictionary in Python for fast lookups) it can be used any number of times, as long as the motif and costs stay the same. Expanding this dictionary for completeness could be another area for further research.

\subsubsection{Multiple Motifs and Time Complexity}

Some forensic DNA loci have multiple motifs (e.\,g., [ACT]5 [AGGT]12). A pair of stutter dictionaries is computed for every motif, and at each step in the RFL algorithm the dictionaries of each motif are checked. This allows for any number and combination of preset motifs to be included in the calculation, each with fast dictionary lookups.

The proposed algorithm allows any prespecified set of motifs. For every additional motif, two inner for-loops are added at each step in the dynamic programming matrix to check the cost of inserting and deleting anywhere from 1 to $2k_i-1$ characters from the dictionary associated with that motif of length $k_i$. Because there are two for-loops running through the list of all motifs, and within each of them there is another for-loop running through $1$ to $2k_i-1$ characters for motif $i$, the algorithm grows in complexity by those factors. Where the classical Levenshtein distance is $O(n_a n_b)$ for strings of length $n_a$ and $n_b$, the RFL distance theoretically requires $O(n_a n_b m k)$ time, where $m$ is the number of motifs and $k$ is the length of the largest motif $\max(k_i)$, assuming $O(1)$ dictionary lookup speed. However, in applications with standard autosomal STRs, motifs are all of length at most 5 ($2k_i-1\leq 9$), so this algorithm can be considered as $O(n_a n_b m),$ growing with the number of motifs considered as well as the lengths of the strings. Additionally, standard autosomal STRs have no more than a handful of motifs to be considered -- a maximum of three, in this work -- so the asymptotic growth becomes $O(n_a n_b)$ once more.

\subsection{Implementation of the RFL Distance}

Consider two strings, $a$ and $b,$ where $|a|=m$ and $|b|=n,$ with zero-based indexing used throughout. The goal of the RFL algorithm is to construct an $(m+1)\times (n+1)$ matrix $d$, where each $d[i,j]$ is equal to the distance from the first $i$ characters of $a$ to the first $j$ characters of $b,$ resulting in the output $d[m,n].$ Recall that the distance is not commutative, so ``parent'' refers to $a$ and ``child'' refers to $b$ (since the string $b$ originally began at $a$ and went through edits to turn into $b$).

Throughout this section, recall the mechanics behind Example \ref{howstackingworks}. The first characters of the parent and child are aligned, and the algorithm then modifies the parent. With each edit the parent is modified to make the leading characters of both strings match, which characters can then be removed for free, and thus the modifications eventually lead to a pair of empty strings and the cost of each step tallied. 

\subsubsection{Building the First Row}

The cost of editing an empty string into an empty string is 0, so $d[0,0]$ is always 0. The element $d[0,1]$ is the cost of editing an empty string into the first character of the child string $b.$ A list of cost options at this step will be made and then the minimal cost option taken. Obtaining the first character can either be done by inserting it directly or by stuttering forward a motif and then deleting characters back. For example, if the first character of the string $b$ was G, and the motifs being considered were GGC and ATTC, then the following costs would be appended to the options list:

\begin{itemize}
    \item The insert cost of the character G
    \item The cost of stuttering forward GGC plus the cost of the weighted Levenshtein distance from GGC to G.
    \item The cost of stuttering forward ATTC plus the cost of the weighted Levenshtein distance from ATTC to G.
\end{itemize}

Each motif can have a different stutter cost associated with it that can vary across loci, and in forensic applications forward stutter will be more expensive than backward since it is a rarer artifact. In this example, GGC might cost 3 to stutter forward and 1 to stutter backward, while ATTC might cost 2 to stutter forward and 0.5 to stutter backward.

The quantity $d[0,1]$ is the minimum of the computed options, and the algorithm moves forward to $d[0,2].$ Now there are several possibilities:

\begin{itemize}
    \item $d[0,1]$ + the cost of inserting the second character via pure insertion (i.e. the cost of inserting $b[1]$, using 0-based indexing)
    \item $d[0,1]$ + the cost of inserting the second character via the stutter dictionary (for each stutter dictionary)
    \item $d[0,0]$ + the cost of inserting the first two characters via the stutter dictionary (for each stutter dictionary)
\end{itemize}

And in general, for each motif, if the motif lengths are $k_i$, for every length $j$ less than or equal to $2k_i-1$, substrings of characters from $j$ positions back to the current can be checked, and if less costly those characters can be inserted via the stutter dictionary. It is impossible to look back further than the beginning of the string, however.

To construct the first row $d[0,:]$, Pythonic pseudocode outlined in Code Excerpt \ref{listing:firstrow} should be followed. Note that \verb|single_char_costs[x,y]| is the cost of editing from character $x$ to character $y$, including the case where $x$ is empty and $y$ is not (corresponding to insertion of $y$), if $y$ is empty and $x$ is not (deletion of $x$), and if $x$ and $y$ are equal (no cost).

\begin{lstlisting}[language=Python,float=htb,caption={Building the first row of the dynamic programming matrix in the RFL algorithm.},captionpos=b,label=listing:firstrow]
lp = len(parent)
lc = len(child)
max_motif_length = max([len(motif) for motif in motifs])
lookahead = 2*max_motif_length - 1
d = zero matrix (lp+1 by lc+1)

for i in 1, ..., lc:
    options = []
    
    options.append(d[0,i-1] + 
        single_char_costs[`',child[i-1]])
    
    for j in 1, ..., min(lookahead,i):
        s = child[i-j:i]
        for motif in motifs:
            if s in costdict[motif][`insert cost']:
                options.append(d[0,i-j] +
                    costdict[motif][`insert cost'][s])
    d[0,i] = min(options)
\end{lstlisting}

Building the first column $d[:,0]$ follows {\em mutatis mutandis}.

\subsubsection{Filling Out the Matrix}

This step combines the complexity of building the first row and column. At each point $d[i,j]$, the algorithm looks back at the previous entries $d[i-1,j], d[i,j-1],$ and $d[i-1,j-1]$ and adds the costs of deletion, insertion, and substitution, respectively. Additionally, for each motif length, it looks back at $d[i-r,j]$ and $d[i,j-r]$, for reverse and forward stutter, respectively, for $r$ from 1 to $2k_i-1$ for each motif $i$ of length $k_i$. After considering all these costs, the entry $d[i,j]$ is the minimum.

In Code Excerpt \ref{listing:fillmatrix}, \texttt{lp} and \texttt{lc} are the lengths of the parent and child strings, respectively. A note on indexing: $d[m-1,n-1]$ is the distance from the first $m-1$ characters of the parent to the first $n-1$ characters of the child. Because of zero-based indexing, parent[$m-1$] is the $m$-th character of the parent.

\begin{lstlisting}[language=Python,float=htb,caption={Building the dynamic programming matrix in the RFL algorithm.},captionpos=b,label=listing:fillmatrix]
for m in 1, ..., lp:
    for n in 1, ..., lc:
        options = []
        
        #substitution
        options.append(d[m-1,n-1] +
            single_char_costs[parent[m-1], child[n-1]])
        
        #insertion
        options.append(d[m,n-1] +
            single_char_costs[`',child[n-1]])
        
        #deletion
        options.append(d[m-1,n] +
            single_char_costs[parent[m-1],`'])
        
        for i in 1, ..., min(lookahead, n):
            s = child[n-i:n]
            for motif in motifs:
                if s in costdict[motif][`insert cost']:
                    options.append(d[m,n-i] + 
                        costdict[motif][`insert cost'][s])
                        
        for i in 1, ..., min(lookahead, m):
            s = parent[m-i:m]
            for motif in motifs:
                if s in costdict[motif][`delete cost']:
                    options.append(d[m-i,n] + 
                        costdict[motif][`delete cost'][s])
        d[m,n] = min(options)
\end{lstlisting}

To demonstrate the RFL algorithm in depth, a fully worked example is given in the supplementary material. 

\subsubsection{A Note on Speed}
The RFL algorithm is written in Python, compatible with NumPy and therefore Numba, which increases the speed over the same algorithm in pure Python by a factor of about 30 via an \texttt{@njit} wrapper. However, the Numba-optimized version of the algorithm only works if including the possibility of stuttering a single motif, due to limitations in the implementation with Numba's inherent dictionary types. Based on testing of the \texttt{weighted-levenshtein} package (written in Cython) \cite{weighted-levenshtein}, the RFL algorithm has potential for significant increases in speed.

\section{Results and Discussion}

\begin{figure}
\centering
\begin{subfigure}{0.5\textwidth}
  \centering
  \includegraphics[width=0.7\linewidth]{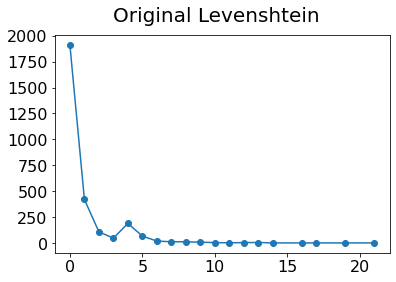}
  \caption{Original Levenshtein.}
  \label{subfig:oglevfreqdist}
\end{subfigure}
\begin{subfigure}{0.5\textwidth}
  \centering
  \includegraphics[width=0.7\linewidth]{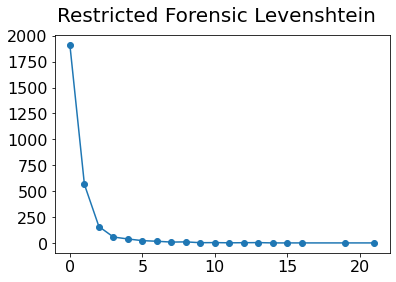}
  \caption{Restricted Forensic Levenshtein.}
  \label{subfig:rflfreqdist}
\end{subfigure}
\caption{Frequency by distance plots at a CSF locus with unit edit costs.}
\label{fig:freqdist}
\end{figure}

An overarching goal of this work was to develop a metric that was inversely proportional to sequence string frequency in MPS data analysis output. A demonstration of the effect of the RFL distance algorithm using the result from Table \ref{tab:Mmt6} is shown in Figure \ref{fig:freqdist}.

\begin{table}
\centering
\begin{tabular}{llrrrlrrr}
   \hline
 Locus & Motif & \makecell{Median \\ LUS} & \makecell{Min \\ LUS} & \makecell{Max \\ LUS} & \makecell{Mean \\ on-LUS \\ back \\ stutter \\ rate} & \makecell{Total \\ parents \\ at locus \\ with \\ motif} & \makecell{Unique \\ parents \\ at locus \\ with \\ motif} \\
   \hline
  CSF1PO & TCTA & 11 & 4 & 15 & 0.0219(163) & 1159 & 17 \\
  D10S1248 & GGAA & 14 & 8 & 19 & 0.0291(190) & 1182 & 13 \\
  D12S391 & TAGA & 12 & 7 & 19 & 0.0279(115) & 1287 & 84 \\
  D12S391 & CAGA & 6 & 3 & 10 & 0.0085(67) & 1287 & 84 \\
  D13S317 & TATC & 12 & 7 & 16 & 0.0183(105) & 1219 & 30 \\
  D13S317 & AATC & 2 & 2 & 3 & 0.0021(56) & 803 & 10 \\
  D16S539 & GATA & 11 & 5 & 14 & 0.0214(123) & 1207 & 17 \\
  D18S51 & AGAA & 15 & 9 & 24 & 0.0249(125) & 1242 & 27 \\
  D19S433 & TCCT & 13 & 7 & 18 & 0.0218(133) & 1194 & 22 \\
  D1S1656 & TATC & 13 & 9 & 17 & 0.0332(141) & 1270 & 30 \\
  D1S1656 & AC & 6 & 5 & 6 & 0.0059(26) & 1270 & 30 \\
  D21S11 & TATC & 12 & 7 & 15 & 0.0224(96) & 1264 & 79 \\
  D21S11 & TGTC & 6 & 4 & 8 & 0.0021(21) & 1264 & 79\\
  D22S1045 & ATT & 12 & 5 & 16 & 0.0316(233) & 1179 & 13 \\
  D2S1338 & GGAA & 13 & 8 & 17 & 0.0221(80) & 1269 & 69 \\
  D2S1338 & GGCA & 7 & 3 & 9 & 0.0033(34) & 1269 & 69 \\
  D2S441 & CTAT & 11 & 7 & 14 & 0.0188(114) & 1205 & 24 \\
  D3S1358 & CTAT & 13 & 5 & 17 & 0.0309(143) & 1238 & 26 \\
  D3S1358 & CTGT & 2 & 2 & 4 & 0.0035(85) & 987 & 19\\
  D5S818 & ATCT & 12 & 7 & 15 & 0.0257(119) & 1237 & 31 \\
  D7S820 & CTAT & 10 & 6 & 13 & 0.0163(96) & 1229 & 30 \\
  D8S1179 & CTAT & 12 & 8 & 15 & 0.0251(115) & 1237 & 34 \\
  D8S1179 & CTGT & 2 & 2 & 3 & 0.0013(51) & 20 & 6\\
  FGA & GAAA & 14 & 9 & 19 & 0.0211(109) & 1247 & 32 \\
  Penta D & GAAAA & 11 & 5 & 17 & 0.0051(47) & 1236 & 18 \\
  Penta E & TTTTC & 12 & 5 & 25 & 0.0114(89) & 1240 & 25 \\
  TH01 & AATG & 7 & 5 & 11 & 0.008(63) & 1165 & 12 \\
  TPOX & AATG & 9 & 5 & 13 & 0.0113(90) & 1140 & 14 \\
  VWA & ATAG & 12 & 3 & 16 & 0.0247(146) & 1242 & 32 \\
  VWA & ACAG & 4 & 3 & 6 & 0.0023(50) & 1242 & 32 \\
  VWA & GATG & 3 & 3 & 4 & 0.0019(8) & 81 & 2\\
   \hline
\end{tabular}
\caption{High-stutter motifs at CODIS-20 and PENTA loci, based on 661 individuals, using a threshold of 0.167 Allele Coverage Ratio (ACR) -- the ratio of the second-highest count to the highest -- to determine homo- vs. heterozygosity in identifying true alleles. For each motif, LUS summary statistics are taken across all alleles. Back stutter rate is the average of all single-stutter proportions resulted from amplification and sequencing, averaged across all loci where parents with that motif were present. The last two columns describe how many alleles and unique alleles with the relevant motif at each locus were present in the data. This analysis is described further in Section \ref{subsec:motifselec}.}
\label{tab:Mmt6}
\end{table}

The non-monotonic plot in Figure \ref{fig:freqdist} has a peak in frequency at a distance of 4 from the true parent sequence in the file, corresponding to stutter of an entire tetranucleotide motif being a more common occurrence than deleting 3 or 5 characters individually. The RFL distance counts this stutter with a cost of 1, thus giving monotonicity. In loci with motifs of other lengths, the frequency peak appears at that particular length. In loci exhibiting multiple motifs, there are increases in frequency at both lengths. There are smaller spikes at a distance of twice the motif length (a distance of 8 in Figure \ref{fig:freqdist}), corresponding to double stutter, but they are exponentially less noticeable.

This metric also has applications in deconvolution, as shown in the following example of real data in a known mixture of two people at the locus D8S1179. According to Table \ref{tab:Mmt6}, the RFL algorithm uses the motifs CTAT and CTGT. Two true alleles at this locus are $p_1$ = [CTAT]12 and $p_2$ = [CTAT]2 CTGT [CTAT]10. The artifact $a$ = [CTAT]2 CTGT [CTAT]9 also appears in the results. It is desirable to capture sequence dissimilarity corresponding to decreased likelihood of an artifact being from a given allele. Using the Levenshtein distance with unit costs, the distance from $p_1$ to $a$ is equal to 1, since the only edit required is changing the third A to a G. The Levenshtein distance from $p_2$ to $a$ is equal to 4, since the motif CTAT must be removed letter-by-letter. However, when using the RFL distance, CTAT can be removed for a cost of 1, so then the distances between the artifact and each true allele are both equal to 1, and the deconvolution is less clear. Future work will fit costs to the data that vary by edit type to more confidently infer which parent was more likely the origin for the artifacts in cases such as this.

\begin{figure}
\centering
\begin{subfigure}{0.475\linewidth}
  \centering
  \includegraphics[width=\textwidth]{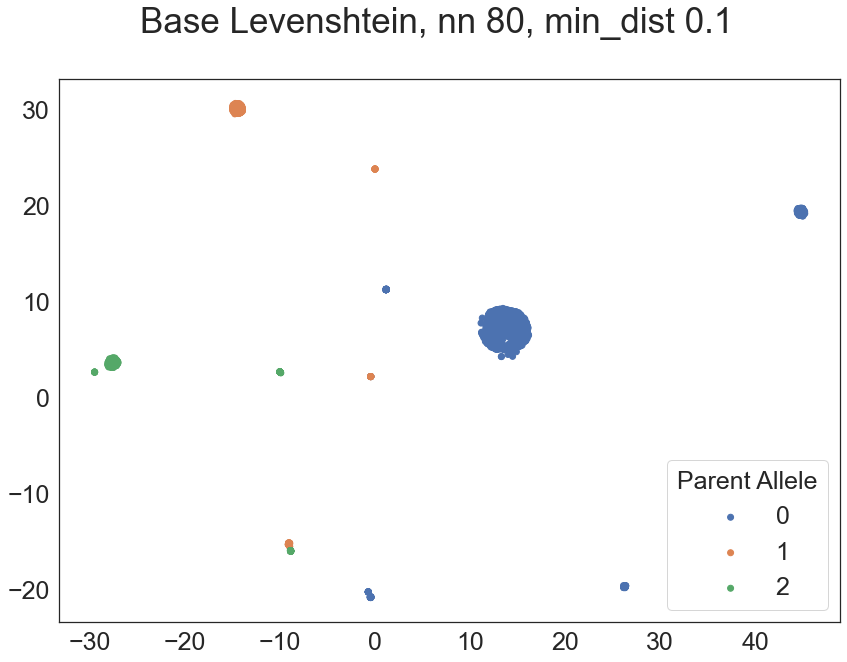}
  \caption{Levenshtein, \texttt{nn} = 80, \texttt{min\_dist} = 0.1.}
  \label{subfig:oglevlow}
\end{subfigure}
\hfill
\begin{subfigure}{0.475\linewidth}
  \centering
  \includegraphics[width=\textwidth]{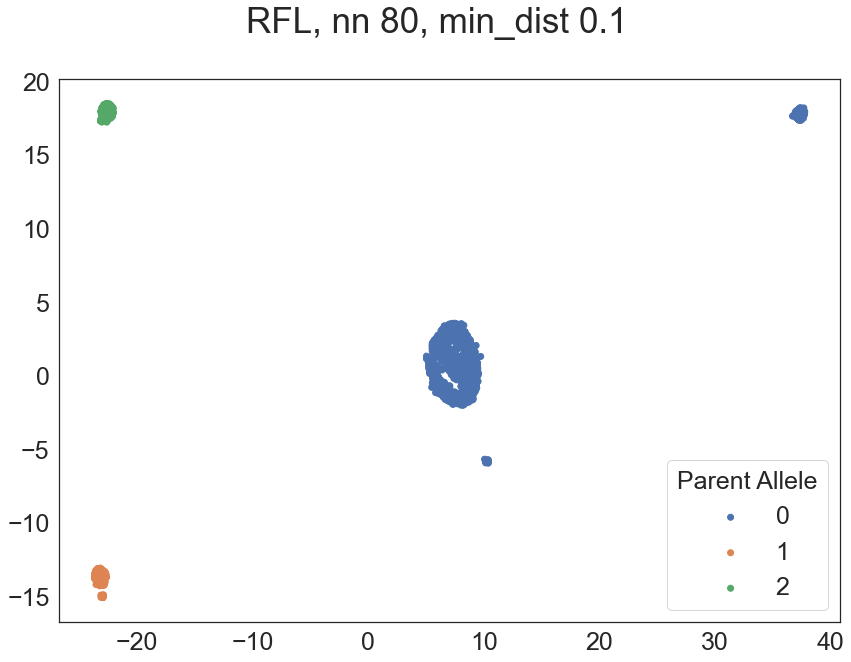}
  \caption{RFL, \texttt{nn} = 80, \texttt{min\_dist} = 0.1.}
  \label{subfig:rfllow}
\end{subfigure}
\caption{UMAP plots for a ground-truth-known mixture of three parts homozygous contributor to one part heterozygous contributor at the locus D1S1656. Each point represents a sequence, and the colors represent which parent that sequence is closest to.}
\label{fig:umap}
\end{figure}

Another way to compare the two metrics is by visualizing the pairwise distances of output in two dimensions. Uniform Manifold Approximation and Projection (UMAP) is one method of performing this dimension reduction technique \cite{umap}. The UMAP algorithm was applied to all pairwise RFL distances and original Levenshtein distances of a 3:1 mixture at locus D1S1656. The locus is typed with three true alleles, with the higher contributor being homozygous and contributing three times as much DNA as the lower contributor, who was heterozygous for this locus. The RFL distance showed more defined clustering than classical Levenshtein across many combinations of hyperparameters, supporting the notion that the RFL distance captures PCR artifact similarity better than original Levenshtein. Figure \ref{fig:umap} shows a plot made with a representative combination of hyperparameters, where each artifact is colored by which parent it is closest to (there were no ties).

\subsection{Selecting Motifs For Each Locus}\label{subsec:motifselec}

The RFL algorithm requires a motif or set of motifs to include as options to lose or gain. The original motivation for the RFL algorithm was to address the high rates of stutter, particularly back stutter, that appear in sequencing data analysis output. In order to give the algorithm the correct motifs to use, the available true allele sequences were exhaustively searched for any motif of length 2-6 that repeated at least twice consecutively. 

For every locus, and for every potential motif identified at that locus, the back stutter rates were computed for all available files by taking the count of the stuttered artifact sequence divided by the sum of all reads for the locus. The results were then filtered to include only loci that showed back stutter at a mean rate of at least 0.001 across samples. Major stutter is the loss of a motif that decreases the longest uninterrupted stretch (LUS) -- i.e., stutter that results in the maximum number of consecutive motifs being reduced. Similarly, minor stutter is the loss of a motif that does not decrease the LUS. This differentiation was made because stutter in a LUS region was more common than stutter outside of a LUS region by several orders of magnitude in the data. Many places in the data showed stretches of a motif at locations other than the LUS region. For example, in every true allele at the locus CSF1PO, a locus with only one STR motif, there were two other locations of that motif repeating in lower numbers than the LUS in the data.

Motifs were grouped by self-permutations -- e.g. if ATAG consecutively repeated at a sequence at least twice, usually TAGA and AGAT would as well. Within each group of self-permutations the motif with the highest mean LUS was chosen, using alphabetization to break ties. The final step was to take the motif at each locus with the largest mean LUS (there were no ties). This final candidate was chosen as the primary motif at the locus -- in essence, the motif that had the highest mean stutter rate at that locus.

Next, every primary motif in every sequence in the data was substituted with the non-nucleotide letter Z, and the entire motif-finding process was repeated to look for secondary motifs -- the motifs that stutter second-most at each locus. This mechanic of replacing the primary motifs with the letter Z ensured there was no overlap of primary and secondary motifs.

This process was iterated until no motifs with high stutter rates and LUS remained. Only VWA had candidate motifs for the tertiary round, and no loci had candidates for quaternary motifs.

Combining the three rounds of processing gives Table \ref{tab:Mmt6}. Within each locus, the rows are sorted in order of the rounds of analysis (and thus in order of stutter frequency).

\section{Conclusion}

A new string dissimilarity measure is proposed with a flexible and publicly available algorithm for implementing it. It has shown promise in ongoing forensic applications thus far, but generalized Levenshtein with multi-character edits could have uses with \textit{in vivo} sequences as well. Applications in graph edit distance and consequent potential uses in handwriting recognition, fingerprint recognition, and cheminformatics were also discussed.

Further work could be done to penalize the overall number of edit operations in addition to the cost of each individual operation. It would be a simple addition to consider transpositions as a separate possible edit operation, though for the original application it was unnecessary.

The restrictions in place because of the stutter dictionary have proven helpful for several reasons, but do limit effectiveness in rare cases. A motif inserting in the middle of another motif, for example, would not be detected by the RFL algorithm, and neither would a motif spread apart by too many single-character insertions ($2\times$len(motif)$-1$ insertions, in particular). Although this limit is easily changeable in the code, the way the algorithm is written requires a hard cap. For forensic applications, however, when the strings are only a few edits apart and insertions are rare, the restrictions do not limit effectiveness.

\section*{Ethics Approval and Consent to Participate}
All work has been reviewed and approved by the National Institute of Standards and Technology Research Protections Office. This study was determined to be ``not human subjects research'' (often referred to as research not involving human subjects) as defined in U. S. Department of Commerce Regulations, 15 CFR 27, also known as the Common Rule (45 CFR 46, Subpart A), for the Protection of Human Subjects by the NIST Human Research Protections Office and therefore not subject to oversight by the NIST Institutional Review Board.

\section*{Data Availability Statement}
    The data that support the findings of this study are from the National Institute of Standards and Technology, but restrictions apply to the availability of these data, which were used under license for the current study, and so are not publicly available. For inquiries, please contact Hari Iyer at hariharan.iyer@nist.gov.

\section*{Competing Interests}
    The authors declare that they have no competing interests.

\section*{Supplementary Material}

The supplementary material for this article explains further complexities that arise when single- and multi-character edits interact, as well as examples computed in detail.

\section*{Funding}
The research of TP and JH was supported in part by the National Science Foundation under Grant No. DMS-1916115 and 2113404. TIH was funded by the NIST Special Programs Office (Forensic Genetics Focus Area).

\section*{Author Contributions}
    
    Conceptualization, T.P., J.H., and H.I.; Methodology, T.P., J.H., and H.I.; Software, T.P.; Validation, T.P., J.H., and H.I.; Formal Analysis, T.P., J.H., and H.I.; Investigation, T.P.; Resources, T.I.H. and H.I.; Data Curation, T.P. and T.I.H.; Writing – Original Draft Preparation, T.P.; Writing – Review \& Editing, T.P., T.I.H., H.I., and J.H.; Visualization, T.P.; Supervision, J.H. and H.I.; Project Administration, J.H. and H.I.; Funding Acquisition, J.H., H.I., and T.I.H.
    
\section*{Acknowledgements}
    Thank you to 
    Katherine B. Gettings, Peter M. Vallone, Lisa A. Borsuk, and the NIST Applied Genetics Group for sharing the data. Special thanks to 
    Katherine B. Gettings for in-depth discussion related to the underlying biological concepts and data details.


\section*{Disclaimer}
    Any commercial equipment, instruments, materials, or software identified in this paper are to foster understanding only. Such identification does not imply recommendation or endorsement by the National Institute of Standards and Technology, nor does it imply that the materials, equipment, or software identified are necessarily the best available for the purpose.

\printbibliography

\pagebreak
\begin{center}
\Large Supplemental Materials:\\ Restricted Forensic Levenshtein Distance
\end{center}
\begin{refsection}
\setcounter{equation}{0}
\setcounter{figure}{0}
\setcounter{table}{0}
\setcounter{page}{1}
\makeatletter
\renewcommand{\theequation}{S\arabic{equation}}
\renewcommand{\thefigure}{S\arabic{figure}}
\renewcommand{\thetable}{S\arabic{table}}

In this supplement, further details of the complexities and examples of the Restricted Forensic Levenshtein algorithm are shown.

\section*{Complexities}

Allowing for the possibility of both single- and multiple-character gains and losses in interaction makes the algorithm exponentially more complex. The insert costs are computed via the weighted Levenshtein distance as if the edit path was first stuttered forward, then edited from the motif (with no stutters) to the string. The delete costs are computed by forcing the string to edit back into a motif, then adding the cost for dropping a motif.

In a case where the motif is AAAGA and all edit costs are equal to 1, evidently the cost of inserting or deleting a single A or G would be 1; however, in the constructed dictionary the cost of these operations are set to 5 each, resulting from a forward stutter (1) and the removal of the remaining 4 letters (4 x 1). The cost of inserting or deleting a letter not part of the main motif, e.g. T, is set to 6 in the dictionary, as it is calculated with the same steps as before (5) and an additional change of the last remaining letter to T (1). Strings of all lengths up to $2k-1$ are also included in the dictionary, i. e. costs of insertion or deletion are set for strings AAA to 3, AAAGAT to 2 and AAACGATC to 4, and so forth.

Cost can be set asymmetrically for insertions and deletions. If insertion costs were changed to 2 but deletions, substitutions, and stutters remained at a cost of 1, the dictionary entries would change. The string AAA in this dictionary would have an insert cost of 3 (blank $\to$ AAAGA (+1) $\to$ AAA (+2)), but a delete cost of 5 (AAA $\to$ AAAGA (+4 for two inserts) $\to$ blank (+1 for stutter)).

It gets more complicated when one considers adding two or three motifs at once, as shown in Example \ref{motifception}.

\begin{example}\label{motifception}
In this example the motif is AAAGA. Forward stutter is followed by many individual letters inserted in between, as shown:
\begin{center}
\textbf{A}(TTTTAAAGATTTT)\textbf{AAGA}.
\end{center}
To get to the same string, another route would be to stutter once and then insert \textit{another motif} in between the letters of the motif that is already there, along with other insertions, as follows:
\begin{center}
\textbf{A}(TTTT)\textbf{AAAGA}(TTTT)\textbf{AAGA}.    
\end{center}
\end{example}

The RFL distance does not include cases where more than $2k-1$ characters fall in between an outer motif's characters.

\section*{A Worked Example}

The example below is shown with unit costs for all operations. The motif is ACG. First an original Levenshtein example is computed in detail, followed by an RFL example. The distance being computed is from ACG to ACGTCG.

Each matrix entry is equal to the distance from the parent string (up to that row) and the child string (up to that column). For example, the bold 4 in Table \ref{standardlevenshteinexample} corresponds to the distance from A to ACGTC. The value at the bottom-right of Table \ref{standardlevenshteinexample} (i.e. 3) is the least costly solution for the edits from parent to child. This follows the intuition that in the base Levenshtein distance, the necessary edits to get from ACG to ACGTCG are three insertions in sequence, each with a cost of 1. The RFL distance between the same two strings with the same motif as in Table \ref{standardlevenshteinexample} is computed in Table \ref{rflexample}.
\begin{center}
    \begin{table}[ht]
        \captionof{table}{Standard Levenshtein} 
        \label{standardlevenshteinexample} 
        \centering
        \begin{tabular}{|c|c|c|c|c|c|c|c|}
            \hline
            \ & \ & \textbf{A} & \textbf{C} & \textbf{G} & \textbf{T} & \textbf{C} & \textbf{G}\\
            \hline
            \ & 0 & 1 & 2 & 3 & 4 & 5 & 6\\
            \hline
            \textbf{A} & 1 & 0 & 1 & 2 & 3 & \textbf{4} & 5\\
            \hline
            \textbf{C} & 2 & 1 & 0 & 1 & 2 & 3 & 4\\
            \hline
            \textbf{G} & 3 & 2 & 1 & 0 & 1 & 2 & 3\\
            \hline
        \end{tabular}
    \end{table}
\end{center}

\begin{center}
    \begin{table}[ht]
        \captionof{table}{Restricted Forensic Levenshtein} 
        \label{rflexample}
        \centering
        \begin{tabular}{|c|c|c|c|c|c|c|c|}
            \hline
            \ & \ & \textbf{A} & \textbf{C} & \textbf{G} & \textbf{T} & \textbf{C} & \textbf{G}\\
            \hline
            \ & 0 & 1 & 2 & 1 & 2 & 3 & 3\\
            \hline
            \textbf{A} & 1 & 0 & 1 & 2 & 2 & 3 & 4\\
            \hline
            \textbf{C} & 2 & 1 & 0 & 1 & 2 & 2 & 3\\
            \hline
            \textbf{G} & 1 & 2 & 1 & 0 & 1 & 2 & 2\\
            \hline
        \end{tabular}
    \end{table}
\end{center}

The following paths can be described from Table 2: A to ACG requires two edits, either via A $\to$ AC $\to$ ACG or A $\to$ AACG $\to$ ACG, while A to ACGTC requires three, via A $\to$ T $\to$ ACGT $\to$ ACGTC, and AC to ACGTCG also requires three, via AC $\to$ ACACG $\to$ ACTCG $\to$ ACGTCG.

The final function value is given in the bottom right corner of the matrix in Table \ref{rflexample}, representing the RFL distance from ACG to ACGTCG. This value of 2 is obtainable via the edit path ACG $\to$ ACGACG $\to$ ACGTCG.

Changing a few individual edit costs will give different optimal paths. In the following example, the cost of inserting a motif (forward stutter) is 2, backward stutter costs 0.5, switching from A to T costs 1.5, switching from A to C costs 0.5, and inserting a C or a T costs 2. All other costs remain equal to 1. The dynamic programming matrix computed with modified costs is shown in Table \ref{rflmodcosts}.

\begin{center}
    \begin{table}[ht]
        \captionof{table}{RFL with modified costs}
        \label{rflmodcosts}
        \centering
        \begin{tabular}{|c|c|c|c|c|c|c|c|}
            \hline
            \ & \ & \textbf{A} & \textbf{C} & \textbf{G} & \textbf{T} & \textbf{C} & \textbf{G}\\
            \hline
            \ & 0 & 1 & 3 & 2.5 & 4.5 & 6.5 & 6.5\\
            \hline
            \textbf{A} & 1 & 0 & 2 & 3 & 4 & 6 & 7\\
            \hline
            \textbf{C} & 1.5 & 1 & 0 & 1 & 3 & 4 & 5\\
            \hline
            \textbf{G} & 0.5 & 1.5 & 1 & 0 & 2 & 4 & 4\\
            \hline
        \end{tabular}
    \end{table}
\end{center}

The previous three examples are shown below, but now with updated costs.

In the previous example, there were two paths from A to ACG that each cost 2. A $\to$ AC (+2) $\to$ ACG (+1) now costs 3, and A $\to$ AACG (+2.5) $\to$ ACG (+1) now costs 3.5. The former path cost matches the cost of 3 in Table \ref{rflmodcosts}, and is thus still a minimal path, but the latter is no longer minimal.

In the previous example, A to ACGTC cost 3. Now the same path A $\to$ T (+1.5) $\to$ ACGT (+2.5) $\to$ ACGTC (+2) costs 6, which matches Table \ref{rflmodcosts}.

For AC to ACGTCG, there was a path that cost 3, but now the cost should be 5 to match Table \ref{rflmodcosts}. AC $\to$ ACACG (+2.5) $\to$ ACTCG (+1.5) $\to$ ACGTCG (+1) costs 5, as desired.

Other cost combinations are possible. For example, if the cost of forward stutter was raised higher than 3 while keeping unit costs for insertion, forward stutter would never be reflected in a minimal edit path because it would always be cheaper to insert the motifs individually instead of stuttering them together.

\section*{An Algorithmic Limitation}

It is possible to weight the costs unequally enough that interferes with the output of the original Levenshtein. If the insert cost of C is 5, but every other edit cost is 1, then the actual cost of inserting a C is 2, since one could insert any other letter (+1) and then substitute in a C (+1). However, the Levenshtein function in the Weighted Levenshtein package available in Python does not reflect this. It will return $lev($`',C$)=5,$ not 2.

Forcing all insert costs to be the same in a given example actually avoids this issue altogether -- for example, the \texttt{stringdist} package in R does not allow for individual insertions to have different costs from one another \cite{r,stringdist}. All deletions must cost the same, as well, as do substitutions.

If the insert cost of C is 1.5, for example, and every other edit cost is 1, then the cost of inserting a C is legitimately 1.5, since any other possible path returns at least 2. In order to have truly minimal-cost answers, then, it is necessary to enter in the individual edit costs in the standard Levenshtein function as minimal (e.g., insertion cost of any letter needs to be less than the sum of any other possible edit sequence to insert that letter indirectly).

Insertion is not the only place where this is a problem. If the cost of A $\to$ T was 5, but all other costs were 1, then A $\to$ C $\to$ T would cost 2, so A $\to$ T should cost 2. A similar example could be constructed for deletion.

The same does not hold true for stutter, however. If the forward stutter cost was raised to 15 and the backward stutter cost raised to 10 in the RFL algorithm, the matrix for the RFL distance (which is equal to the standard Levenshtein matrix) is shown in Table \ref{rfl_coststoohigh}.

There are no costs of 10 or higher, since it is never cheaper to stutter than to address each nucleotide separately. Thus, the addition of motifs to the Weighted Levenshtein package does not suffer from the same weaknesses. Because it uses the package as a base, though, it still has the same problems as the original Levenshtein distance with the other edit operations.

\begin{center}
    \begin{table}[h]
        \captionof{table}{RFL with high stutter penalties}
        \label{rfl_coststoohigh} 
        \centering
        \begin{tabular}{|c|c|c|c|c|c|c|c|}
            \hline
            \ & \ & \textbf{A} & \textbf{C} & \textbf{G} & \textbf{T} & \textbf{C} & \textbf{G}\\
            \hline
            \ & 0 & 1 & 2 & 3 & 4 & 5 & 6\\
            \hline
            \textbf{A} & 1 & 0 & 1 & 2 & 3 & 4 & 5\\
            \hline
            \textbf{C} & 1 & 1 & 0 & 1 & 2 & 3 & 4\\
            \hline
            \textbf{G} & 3 & 2 & 1 & 0 & 1 & 2 & 3\\
            \hline
        \end{tabular}
    \end{table}
\end{center}

The solution to this issue is a pre-processing step to ensure that the individual edit costs are already optimal prior to the application of the Levenshtein distance. It is more complicated than, e.g., ensuring that a two-step insert-substitute path is not cheaper than direct insertion. Pathological counterexamples exist that require more than two steps -- a demonstration is shown in Example \ref{pathological}.

\begin{example}\label{pathological}
In a case where the following costs are set:
\begin{itemize}
    \item Cost to insert a C, T, or G: 10
    \item Cost to insert an A: 1
    \item Cost of T $\to$ C, A $\to$ C, or A $\to$ G: 10
    \item Cost of G $\to$ C, A $\to$ T, or T $\to$ G: 1
\end{itemize}

Then in order to insert a C, it is cheaper to edit `' $\to$ A $\to$ T $\to$ G $\to$ C for a total cost of 4 than it is to insert a C directly (10), but one cannot go `' $\to$ A $\to$ C (cost of 11) or `' $\to$ A $\to$ T $\to$ C (cost of 12).
\end{example}

In the application the RFL was designed for, however, all deletions cost the same, all insertions cost the same, and all substitutions cost the same, so this issue is avoided.

The RFL algorithm is based on the idea that the sequences undergoing the edits in PCR should take the cheapest, most frequent path. Here distance is defined to be the lowest possible cost of any edit path between the strings (regardless of whether the minimal path is itself unique), with no penalty on the number of steps. The implementation of edit count penalization to the algorithm is an rea of interest for future work.

\printbibliography
\end{refsection}
\end{document}